\documentclass[9pt,conference]{IEEEtran}
\IEEEoverridecommandlockouts

\usepackage{cite}
\usepackage{amsmath,amssymb,amsfonts}
\usepackage{graphicx}
\usepackage{textcomp}

\usepackage{hyperref}
\usepackage{colortbl}
\usepackage{multirow}
\usepackage[table,xcdraw]{xcolor}
\usepackage{booktabs}
\usepackage{algorithm}
\usepackage{algpseudocode}
\usepackage{amsmath}
\usepackage{amsfonts}
\usepackage{adjustbox}
\usepackage{graphicx}
\usepackage{subcaption}
\usepackage{multirow}
\usepackage{colortbl}
\usepackage{hhline}
\usepackage{forest}
\usepackage{adjustbox}
\usepackage{colortbl}
\usepackage[table,xcdraw]{xcolor}
\hypersetup{
    colorlinks,
    citecolor=green,
    filecolor=black,
    linkcolor=blue,
    urlcolor=black
}

\usepackage{array}
\usepackage{multirow}
\usepackage{ragged2e}
\usepackage{hhline}
\usepackage{colortbl}

\def\BibTeX{{\rm B\kern-.05em{\sc i\kern-.025em b}\kern-.08em
    T\kern-.1667em\lower.7ex\hbox{E}\kern-.125emX}}
\begin{document}
\title{An Explainable Probabilistic Attribute Embedding Approach for Spoofed Speech Characterization\\
\thanks{The work has been partially supported by the Academy of Finland (Decision No. 349605, project "SPEECHFAKES"). The authors wish to acknowledge
CSC – IT Center for Science, Finland, for computational resources.}
}

\author{\IEEEauthorblockN{Manasi Chhibber$^1$, Jagabandhu Mishra$^1$, Hyejin Shim$^2$, and Tomi H. Kinnunen$^1$
}
\IEEEauthorblockA{$^1$University of Eastern Finland, Finland \\
$^2$Carnegie Mellon University, USA
}}

\maketitle

\begin{abstract}
We propose a novel approach for spoofed speech characterization through explainable probabilistic attribute embeddings. In contrast to high-dimensional raw embeddings extracted from a spoofing countermeasure (CM) whose dimensions are not easy to interpret, the probabilistic attributes are designed to gauge the presence or absence of sub-components that make up a specific spoofing attack. 
These attributes are then applied to two downstream tasks: spoofing detection and attack attribution. To enforce interpretability also to the back-end, we adopt a decision tree classifier. Our experiments 
on the ASVspoof2019 dataset with spoof CM 
embeddings extracted from three models (AASIST, Rawboost-AASIST, SSL-AASIST) 
suggest that the performance of the 
attribute embeddings are on par with the original raw spoof CM embeddings for both tasks. The best performance achieved with 
the proposed approach 
for spoofing detection and attack attribution, in terms of accuracy, is $99.7\%$ and $99.2\%$, respectively, compared to $99.7\%$ and $94.7\%$ using the raw CM embeddings. To analyze the relative contribution of each attribute, we estimate their Shapley values. Attributes related to acoustic feature prediction, waveform generation (vocoder), and speaker modeling are found important for spoofing detection; while duration modeling, vocoder, and input type play a role in spoofing attack attribution.
\end{abstract}



\section{Introduction}
Modern speech synthesis systems produce increasingly human-like speech in a targeted person's voice~\cite{shen2018natural}. 
While offering numerous benefits, these developments also 
increase security and privacy risks due to the potential for misuse, such as spoofing voice biometric systems~\cite{todisco19_interspeech,lan2022adversarial,jung22c_interspeech}. To mitigate these risks, the research community has proposed various \emph{countermeasures} (CMs), typically in the form of a binary (bonafide human vs. spoof) classifier~\cite{todisco19_interspeech,lan2022adversarial,jung22c_interspeech}. However, as these solutions are typically developed with large-scale, fully automated applications in mind, their functioning can be nontransparent to humans. This is due to both the large number of parameters and lack of providing explanations for model predictions. 

While explainability may not be crucial for all applications, it is essential in critical areas such as forensics~\cite{amor2022ba}. Very few studies, however, have addressed the timely need for interpretability in antispoofing.
The prior studies 
broadly follow two ways. The first one utilizes visualization techniques 
such as \emph{gradient-weighted class activation mapping} (Grad-CAM)\cite{liu24m_interspeech} and \emph{saliency maps}\cite{gupta24b_interspeech} to highlight regions of a speech waveform or spectrogram that contribute to the prediction. Although these approaches provide some level of explanation, they do not directly relate to the process of spoofed speech generation.



The second approach is to explain spoofed speech through its generation process by identifying the source. For example, spoofed speech might be generated using a specific text-to-speech (TTS) or voice conversion (VC) method. The task of identifying the source of spoofed speech, or \emph{spoofing attack attribution}~\cite{bhagtani2023synthesized}, 
has been addressed using features such as bispectral magnitude and phase~\cite{albadawy2019detecting}, loudness, shimmer, spectral energy~\cite{muller2022attacker}; and data-driven techniques ranging from self-supervised learning~\cite{yadav2023synthetic} and convolutional neural networks (CNNs)~\cite{salvi2022exploring} to patchout transformers~\cite{bhagtani2023synthesized}. Whereas spoofing attack attribution aims at identifying the complete spoofing method, another, more fine-grained version of the task seeks to identify the components that make up a specific attack. For instance, modern TTS and VC systems use a cascade of \emph{acoustic model} (e.g., Tacotron~\cite{yasuda2019investigation} or FASTPITCH~\cite{lancucki2021fastpitch}) and \emph{vocoder} (e.g., Wavenet~\cite{oord2018parallel} or MelGAN~\cite{kumar2019melgan}) to generate or modify speech. With this motivation, the authors of~\cite{klein24_interspeech} attempt to identify (termed "source tracing") these two components, along with the input type (text/speech). Although these approaches 
are designed to identify the spoofing attacks or their components, similar to standard CMs, they are primarily designed with predictive performance---rather than interpretability---in mind. 


\begin{figure}[t!]
    \centering
     \fbox{\includegraphics[height= 120pt,width=240pt]{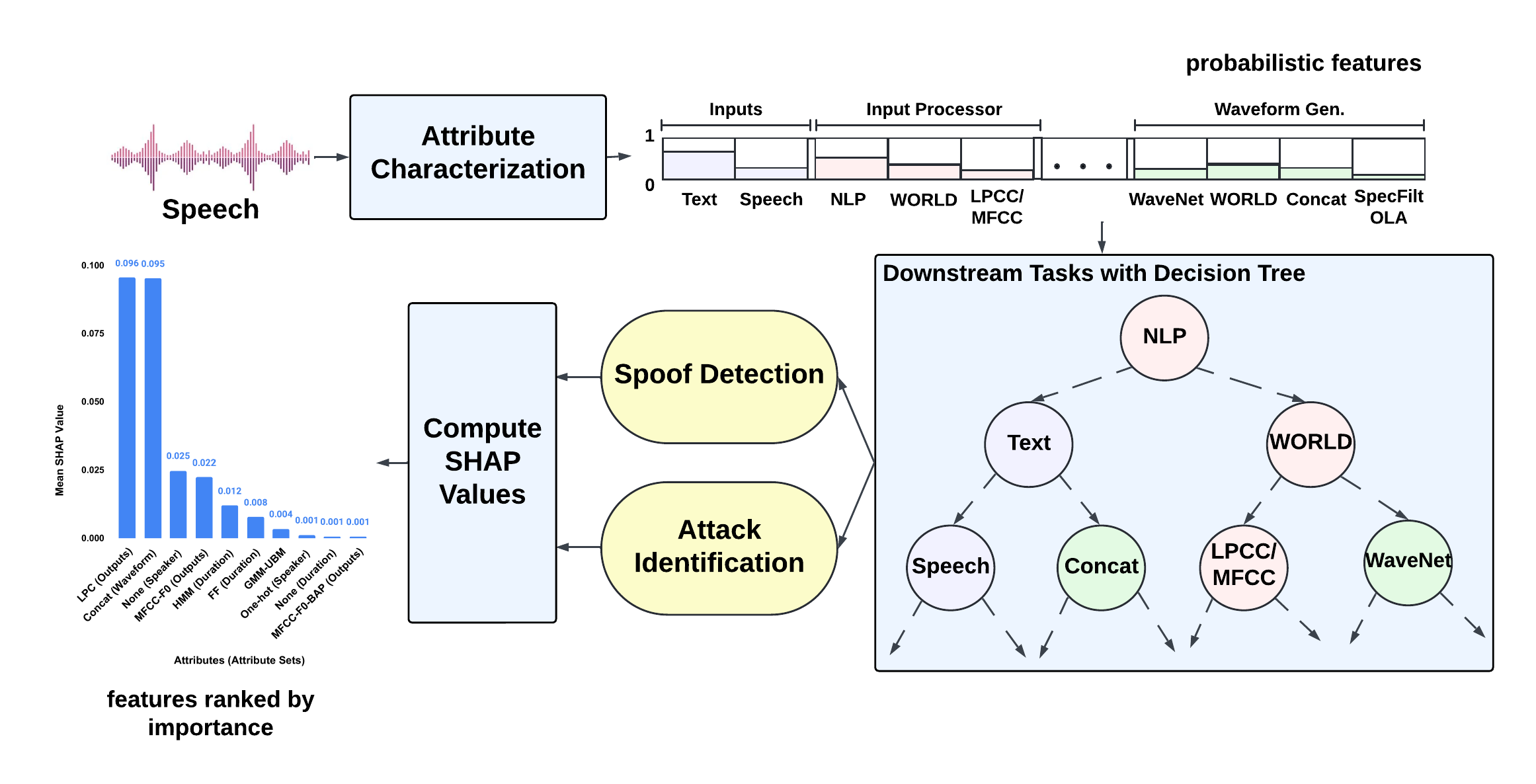}}
    \caption{Block diagram of the proposed {\it probabilistic attributes} based explainable framework. }
    \label{intro_fig}
\vspace{-0.5cm}
\end{figure}

To address the need for more explainable spoofed speech characterization, we propose a novel \textit{probabilistic attribution} framework. Unlike existing approaches that rely on opaque CMs offering limited insight into their decision-making, our framework emphasizes “explainability-by-design.” We achieve this by introducing interpretable probabilistic attributes, defined based on the metadata of speech generation algorithms. 
These attributes are then used for both spoofing detection and spoofing attack attribution, using an interpretable decision tree classifier ~\cite{song2015decision}. While previous studies have used Shapley values ~\cite{scott2017unified} to analyze the contributed regions in the speech signal and its spectrogram for spoofing detection ~\cite{ge2022explaining}, our work extends this approach by analyzing the relative contribution of each probabilistic attribute for both spoofing detection and spoofing attack attribution.

\begin{figure*}[]
    \centering
    \includegraphics[height= 180pt,width=460pt]{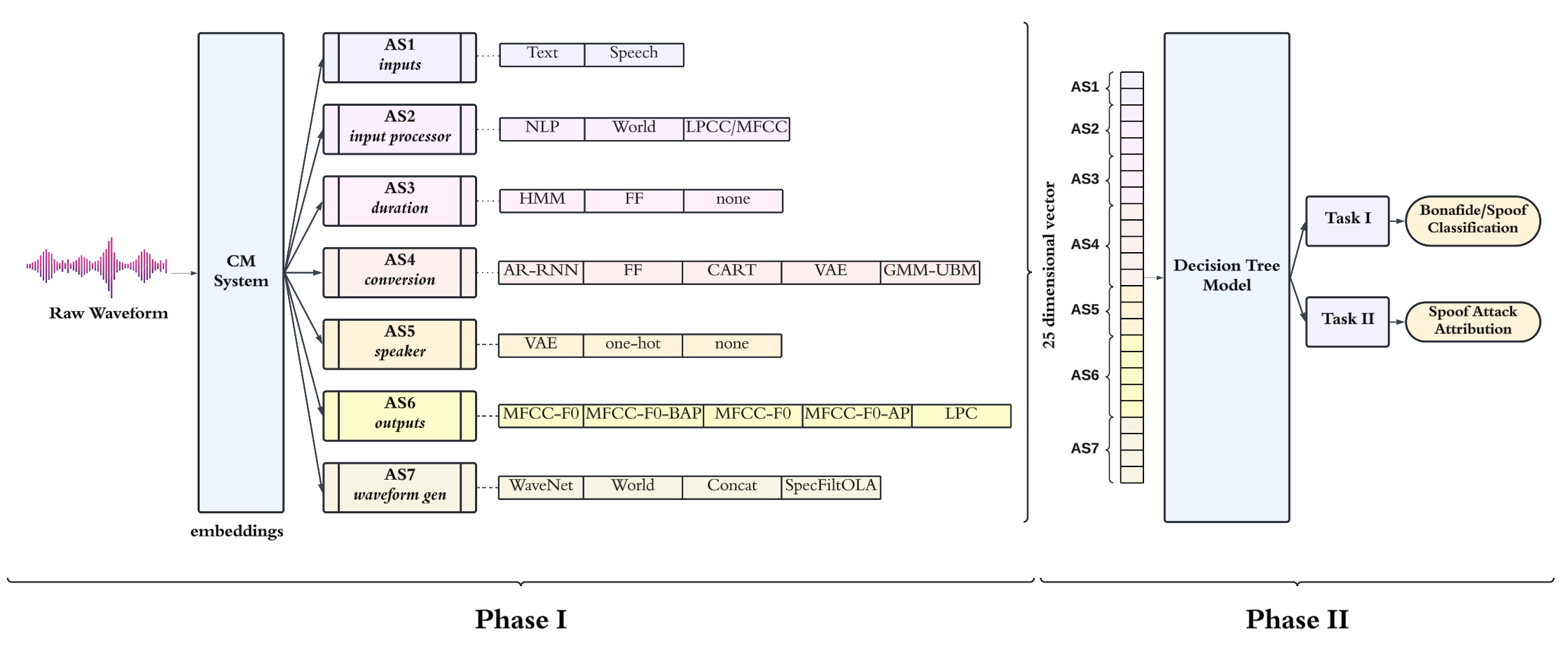}
    \caption{Overall pipeline of the proposed architecture for explainable spoofed speech detection. \textbf{Phase I} demonstrates the extraction of embeddings using a countermeasure system and the subsequent processing of these embeddings through a bank of seven probabilistic feature detectors. \textbf{Phase II} illustrates the concatenation of the outputs from these detectors to create a $25$-dimensional vector, which is then fed into a decision tree model for classification. This decision tree model is used for both bonafide/spoofed classification and spoofing attack attribution.}
    \label{fig:full_arch}
    \vspace{-0.3cm}
\end{figure*}

\section{Downstream tasks}
We consider two downstream tasks: (1) spoofing detection (binary) and (2) spoofing attack attribution (multi-class). In contrast to ``raw" high-dimensional spoofing CM embeddings, we address both tasks by using low-dimensional, interpretable probabilistic attribute embeddings and a decision tree classifier. The Shapley approach is then used to explain the importance of each attribute in performing either task. The overall workflow is illustrated in Fig.~\ref{fig:full_arch}. A brief description of each approach is provided in the following subsections. 

\subsection{Spoofing Detection}

Spoofing detection~\cite{todisco19_interspeech} involves classifying a given utterance as either bonafide or spoofed. Given a speech utterance ${ \bf x}$, the goal is to train a classifier ($\mathcal{F}^{\text{cm}}$) to predict either the null hypothesis ($\mathcal{H}_0$, bonafide) or the alternate hypothesis ($\mathcal{H}_1$, spoof), expressed as $\mathcal{F}^{\text{cm}}:{\bf x} \mapsto \{\mathcal{H}_0,\mathcal{H}_1\}$. This task is typically addressed using an end-to-end deep learning model~\cite{todisco19_interspeech}. However, in this work, for the sake of comparison with the proposed explainable probabilistic attribute embeddings (${\bf p}_a$), we extract spoof CM embeddings (${\bf e}_{\text{cm}}$) from the trained classifier and use them as the input to an interpretable decision tree model. Specifically, for a given speech utterance ($\bf x$), the spoof CM embedding is extracted as, $\mathcal{F}^{\text{cm}}:{\bf x} \mapsto {\bf e}_{\text{cm}}$.  The baseline approach is represented as $\mathcal{F}^{\text{DT}}: {\bf e}_{\text{cm}} \mapsto \{\mathcal{H}_0,\mathcal{H}_1\}$, while the proposed approach is represented as $\mathcal{F}^{\text{DT}}: {\bf p}_a \mapsto \{\mathcal{H}_0,\mathcal{H}_1\}$.


\subsection{Spoofing Attack Attribution}
While spoofing detection is a binary classification task, spoofing attack attribution~\cite{bhagtani2023synthesized} aims to predict  
 the type of attack used to generate a given speech utterance (${\bf x}$). Suppose that spoofed utterances can originate from $N$ known spoofing attacks,  $\mathcal{A}=\{\mathcal{A}_1,\dots,\mathcal{A}_N\}$. The spoofing attack attribution task is expressed as $\mathcal{F}^{\text{saa}}:{\bf x} \mapsto \{\mathcal{A}_1,\dots,\mathcal{A}_N\}$. This task has been approached both through end-to-end deep learning methods~\cite{bhagtani2023synthesized} or by using intermediate CM embeddings ${\bf e}_{\text{cm}}$ extracted from $\mathcal{F}^{\text{cm}}$, followed by simple fully connected neural network classifier~\cite{klein24_interspeech}. Similar to spoofing detection, as a baseline, we use embeddings extracted from the CM system with a decision tree classifier to perform spoofing attack attribution,  $\mathcal{F}^{\text{DT}}:{\bf e_{\text{cm}}} \mapsto \{\mathcal{A}_1,\dots,\mathcal{A}_N\}$. The proposed approach, using probabilistic attribute embeddings, performs spoofing attack attribution with a decision tree classifier and is represented as $\mathcal{F}^{\text{DT}}:{\bf p}_a \mapsto \{\mathcal{A}_1,\dots,\mathcal{A}_N\}$.


\section{Attribute Characterization of Spoofed Speech and its Explanability}
\subsection{Attribute Characterization of Spoofed Speech}
The VC and TTS algorithms are typically designed either through cascaded modules such as intermediate acoustic feature predictor followed by a vocoder; or via end-to-end approaches~\cite{salvi2023timit}. In this initial study,  we focus exclusively on the former, as the aim is to extract information on the presence or absence of the modules. The acoustic feature predictor is further characterized by input type (text/speech), preprocessing methods, duration modeling, acoustic features, etc. Combining all of these modules, we define {\it probabilistic attribute} embedding as, ${\bf p}_a= ({\bf a}_{1},\ldots,{\bf a}_L)$, where ${\bf a}_i$ represents an attribute set specific to a module, and $L$ is the number of modules considered. The particular method used within each module is defined as an attribute ($a_{j}$). Thus, a specific attribute set is defined as ${\bf a}_{i}=(a_1,\ldots, a_{M_i})$, where $a_j \in (0,1)$, and $M_i$ is the number of possible methods within a specific module.



 
 In this study, for a given speech utterance ($\bf x$), we first extract a spoof CM embedding (${\bf e}_{\text{cm}}$) using a suitable CM system ($\mathcal{F}^{\text{cm}}$). The embeddings are then used to train $L$ attribute classifier networks ($\mathcal{F}^{\text{ac}}_{{\bf a}_{i}}$), each specific to an attribute set as follows,

 \begin{equation}
     \mathcal{F}^{\text{ac}}_{{\bf a}_{i}}:{\bf e}_{\text{cm}} \mapsto {\bf a}_{i} \in \mathbf{P}^{M_i},
 \end{equation}
where $1 \leq i \leq L$, and $\mathbf{P}$ is probability simplex, i.e. $\mathbf{P}^M=\{(p_1,\ldots,p_M):p_m \geq 0,\sum_{m=1}^M p_m=1\}$. Finally, the probabilistic attribute embedding is represented as the concatenated vector of the attribute vectors extracted from each module, i.e. ${\bf p}_a=(a_{j}^{i})$, and can also be represented as ${\bf p}_a=(a_{k})$, $1 \leq k \leq T$, and $T=\sum_i^L M_i$  is the total number of attributes in the probabilistic attribute embedding.

\subsection{Decision Trees and Shapley values}
\emph{Decision tree} \cite{song2015decision},  a classic hierarchical model specified through a set of simple feature thresholding rules, is known for its inherent interpretability   \cite{freitas2014comprehensible}. 
The decision tree is fed with probabilistic attribute vectors, where each node represents an attribute (\(a_k\)). For a given classification task, the tree structure helps identify the most prominent features by positioning the significant attribute nodes closer to the tree root. However, the tree alone does not provide a measurable score for feature importance. To address this and quantify the contribution of each attribute towards the prediction, we use \emph{Shapley values}~\cite{scott2017unified}, which are widely employed in machine learning literature.



\emph{Shapley values} measure the average marginal contribution of a feature to the prediction across all possible combinations of features. Mathematically, the Shapley value of an attribute (\(a_k\)) of probabilistic attributes vectors (${\bf p}_a$) using the decision tree classifier ($\mathcal{F}^{DT}$) is calculated as:

\begin{equation}
\begin{aligned}
\phi_k&(\mathcal{F}^{DT},{\bf p}_a ) =\\
                                    &\sum_{S \subseteq {\bf p}_a \setminus {a_k}} \frac{|S|!(T - |S| - 1)!}{T!} [\mathcal{F}^{DT}(S \cup {a_k}) - \mathcal{F}^{DT}(S)],
\end{aligned}
\end{equation}

\noindent
where \(S\) represents a subset of features excluding \(a_k\), and \(|S|\) denotes the cardinality of \(S\). In practice, these Shapley values are computed from the decision tree by traversing through all possible paths from the root node to a leaf node. For each path, we evaluate the prediction based on the combination of features present in that path, including or excluding the attribute \(a_k\). The difference between these predictions, representing the marginal contribution of \(a_k\), is then calculated. This process is repeated for all possible paths, and the Shapley value for \(a_k\) is obtained by averaging these marginal contributions. In this study, we calculate the Shapley values of attributes \(a_k\) to understand their contributions in performing the downstream tasks. This analysis will help us to identify the key decision-makers in the classification process and quantify their individual contributions to the downstream tasks.

\section{Experimental Setup, Results and Discussion}
In this work, we considered three spoofing CM approaches: Audio Anti-Spoofing using Integrated Spectro-Temporal Graph Attention Networks (AASIST)~\cite{jung2022aasist}, Rawboost-AASIST~(RB-AASIST)\cite{tak2022rawboost}, and Self-Supervised AASIST (SSL-AASIST)~\cite{tak2022automatic} for characterizing the probabilistic attribute embeddings.
The AASIST model, initially developed for binary classification, was chosen for adaptation due to its wide usage and state-of-the-art performance. To address the risk of overfitting to known speaker attributes in the training set, we integrated Rawboost and self-supervised learning (SSL) techniques into the AASIST model, enhancing its generalization capabilities.
The equal error rate (EER) is used as the evaluation metric for evaluating the performance of each attribute classifier network ($\mathcal{F}^{\text{ac}}_{{\bf a}_{i}}$). Although prior studies evaluate spoofing detection using EER, the decision tree classifier used in this study does not provide probability prediction scores. Therefore, both spoofing detection and spoofing attack attribution are evaluated using accuracy and F1 score. The implications of the methods used are available on GitHub for reference~\footnote{\href{https://github.com/Manasi2001/Spoofed-Speech-Attribution}{https://github.com/Manasi2001/Spoofed-Speech-Attribution}.}.




\subsection{Database}
For all experiments in this paper, we use the ASVspoof2019 logical access (LA) dataset. Compared to other datasets in the literature, the clean, high-quality recordings of ASVspoof allow us to focus exclusively on interpretability, which is the primary motivation for using this dataset. The ASVspoof 2019 LA dataset is divided into training, development, and evaluation parts. The training and development sets contain both bonafide speech and spoofed speech generated using $6$ different TTS and VC systems that are designated as known attacks. 
The evaluation set includes bonafide and spoofed utterances generated by $13$ TTS and VC systems, of which $2$ are known and $11$ are unknown spoofing attacks. The attribute vectors are defined, and attribute extractors are trained using the training set. Model selection and tuning hyperparameter for max depth in the decision tree are performed on the development set. The evaluation and interpretation of results for both downstream tasks are carried out on both the development and evaluation sets. However, for spoofing attack attribution, the performance is evaluated only in the known attack types present in the evaluation set.

\vspace{-0.1cm}
\subsection{Phase I: Attribute Characterization}

\begin{figure}[t!]
    \centering
    \includegraphics[height= 120pt,width=240pt]{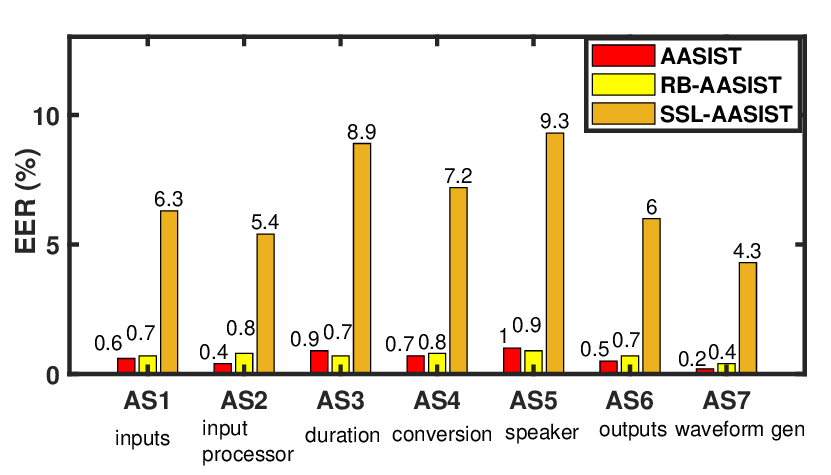}
    \caption{Attribute characterization results of each attribute set (AS) shown in terms of EER.  }
    \label{fig:as_results}
    \vspace{-0.4cm}
\end{figure}

\begin{figure}[t!]
    \centering
    {\includegraphics[height= 120pt,width=240pt]{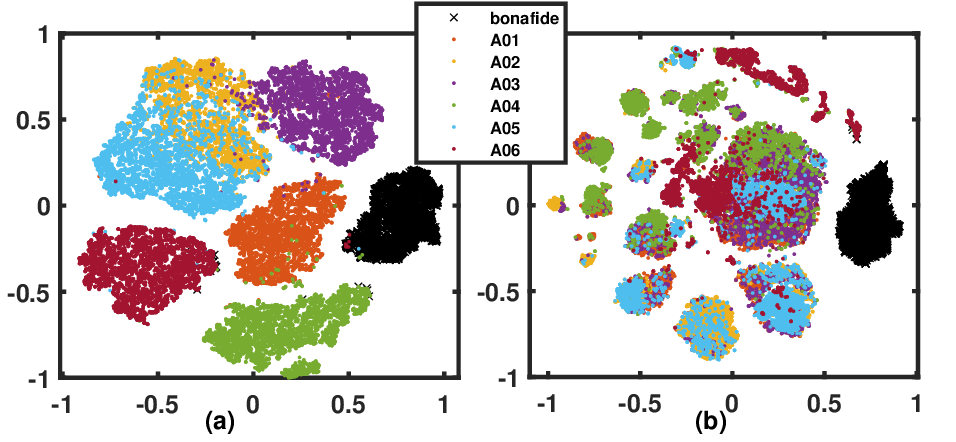}}
    \caption{The t-SNE projection of spoof CM embeddings obtained from (a) AASIST, and (b) SSL-AASIST systems, respectively.}
    \label{tsne_1}
    \vspace{-0.5cm}
\end{figure}

The metadata information in \cite[Table~1]{wang2020asvspoof}, suggests that seven distinct modules are used to generate the spoofed utterances. An attack type $\mathcal{A}$ is defined by choosing a specific method from each module. In this study, the distinct methods used in each module within the training and development sets are treated as attributes ($a_k$), and the modules themselves are considered as the attribute set (${\bf a}_i$)  for characterizing the probabilistic attribute embeddings. The total number of attributes across all the attribute sets is $25$. The considered attribute set and attributes are defined in Fig.~\ref{fig:full_arch}. For each spoofed utterance specific to a module, the ground truth attribute vector is represented as a binary vector, where a value of $1$ is assigned if the corresponding method was used in generating the spoofed utterance. The ground truth binary vectors are used as training targets to train the attribute classifier networks. Each attribute classifier network takes $160$ dimensional spoof CM embeddings as input and provides the number of attributes in that attribute set $M_i$ as output. Each classifier has two hidden nodes of $64$ and $32$ neurons, respectively. Output neurons used softmax, and all others used rectified linear unit (Relu) activation function. Each classifier network is trained for $100$ epochs using the Adam optimizer with an adaptive learning rate, starting with a base learning rate of $0.0001$. The epoch corresponding to the least EER in the development set is used for the probabilistic attribute embedding extraction.  The architecture and training protocol is finalized after running a few experiments and observing the performance in the development set. The obtained EER in each attribute set with three spoof CM embeddings are shown in Fig.~\ref{fig:as_results}. 

It is worth noting that the spoofing detection performance, in terms of EER, for ASSIST~\cite{jung2022aasist}, RB-ASSIST~\cite{tak2022rawboost}, and SSL-AASIST~\cite{tak2022automatic} on the evaluation set is $0.83\%$, $1.59\%$, and $0.22\%$, respectively. These results indicate that SSL-AASIST is a more effective spoofing CM classifier than AASIST. However, the results in Fig.~\ref{fig:as_results} suggest that, regardless of the attribute set, AASIST characterizes the attributes more effectively than SSL-AASIST. This may be due to the SSL model being pre-trained on a different dataset, which could lead to generalization across attack types during fine-tuning for bonafide/spoof classification. In contrast, AASIST is trained directly on the ASVspoof dataset, hence the CM embeddings are likely to retain traces of specific attack types. To validate this hypothesis, the CM embeddings from both AASIST and SSL-AASIST were projected into two dimensions using t-SNE~\cite{van2008visualizing}, and the resulting representations are shown in Fig.~\ref{tsne_1}. The figure demonstrates that while bonafide examples form distinct clusters in both embedding spaces, attack types are more overlapped in the SSL-AASIST space compared to AASIST. This observation suggests that spoof attack attribution may be more effective using AASIST embeddings, while spoof detection performance may be better with SSL-AASIST embeddings.

\subsection{Phase II: Downstream Task with Probabilistic Attributes}

The spoof CM embeddings and the probabilistic attribute embeddings are used with the decision tree classifier to perform the spoofing detection and spoofing attack attribution task. The obtained performances of spoofing detection and spoofing attack attribution are tabulated in Table~\ref{SDP} and~\ref{tab2_new}.

As expected, the results suggest that the embeddings ${\bf e}_{\text{cm}}$ and ${\bf p}_a$ extracted from SSL-AASIST perform better in spoofing detection, while the same embeddings from AASIST are more effective in spoofing attack attribution. Additionally, regardless of the CM system, the performance of ${\bf p}_a$ in spoofing detection is on par with ${\bf e}_{\text{cm}}$, and superior in spoofing attack attribution. This demonstrates that by representing the CM embeddings through ${\bf p}_a$, we do not lose performance in downstream tasks. Furthermore, we gain better explainability, as the constructed decision tree can meaningfully explain the decision-making process.  However, while the decision tree provides interpretability, it does not quantify feature importance. Therefore, in the following section, we discuss this aspect using the Shapley values.

\begin{table}
\centering
\caption{Performance of \emph{spoofing detection} with spoof CM embedding (${\bf e}_{\text{cm}}$) and probabilistic attribute embedding (${\bf p}_a$) extracted from AASIST, RB-AASIST and SSL-AASIST systems.}
\label{SDP}
\arrayrulecolor{black}
\begin{tabular}{|c|c|c|c|c|c|c|c|} 
\hline
\rowcolor[rgb]{0.925,0.957,1} \multicolumn{2}{|c|}{{\cellcolor[rgb]{0.925,0.957,1}}}                                                               & \multicolumn{6}{c|}{\textbf{Spoofing Detection~}}                                                                                                                   \\ 
\hhline{|>{\arrayrulecolor[rgb]{0.925,0.957,1}}-->{\arrayrulecolor{black}}------|}
\rowcolor[rgb]{0.925,0.957,1} \multicolumn{2}{|c|}{{\cellcolor[rgb]{0.925,0.957,1}}}                                                               & \multicolumn{2}{c|}{\textbf{AASIST~}}                & \multicolumn{2}{c|}{\textbf{RB-AASIST~}}             & \multicolumn{2}{c|}{\textbf{SSL-AASIST~}}             \\ 
\hhline{|>{\arrayrulecolor[rgb]{0.925,0.957,1}}-->{\arrayrulecolor{black}}------|}
\rowcolor[rgb]{0.882,0.941,0.882} \multicolumn{2}{|c|}{\multirow{-3}{*}{{\cellcolor[rgb]{0.925,0.957,1}}~}}                                        & ${\bf e}_{\text{cm}}$ & {\cellcolor[rgb]{1,1,0.78}}${\bf p}_a$ & ${\bf e}_{\text{cm}}$ & {\cellcolor[rgb]{1,1,0.78}}${\bf p}_a$ & ${\bf e}_{\text{cm}}$ & {\cellcolor[rgb]{1,1,0.78}}${\bf p}_a$  \\ 
\hline
\rowcolor[rgb]{0.882,0.941,0.882} {\cellcolor[rgb]{0.925,0.957,1}}                                 & {\cellcolor[rgb]{0.925,0.957,1}}\textbf{Acc~} & 99.6~       & {\cellcolor[rgb]{1,1,0.78}}99.7~       & 99.3~       & {\cellcolor[rgb]{1,1,0.78}}99.0~       & \textbf{99.9}~       & {\cellcolor[rgb]{1,1,0.78}}\textbf{99.9}~        \\ 
\hhline{|>{\arrayrulecolor[rgb]{0.925,0.957,1}}->{\arrayrulecolor{black}}-------|}
\rowcolor[rgb]{0.882,0.941,0.882} \multirow{-2}{*}{{\cellcolor[rgb]{0.925,0.957,1}}\textbf{Dev~}}  & {\cellcolor[rgb]{0.925,0.957,1}}\textbf{F1~}  & 0.99~       & {\cellcolor[rgb]{1,1,0.78}}0.99~       & 0.98~       & {\cellcolor[rgb]{1,1,0.78}}0.99~       & 1.00~       & {\cellcolor[rgb]{1,1,0.78}}1.00~        \\ 
\hline
\rowcolor[rgb]{0.882,0.941,0.882} {\cellcolor[rgb]{0.925,0.957,1}}                                 & {\cellcolor[rgb]{0.925,0.957,1}}\textbf{Acc~} & 99.4~       & {\cellcolor[rgb]{1,1,0.78}}98.8~       & 98.6~       & {\cellcolor[rgb]{1,1,0.78}}98.9~       & \textbf{99.7}~       & {\cellcolor[rgb]{1,1,0.78}}\textbf{99.7}~        \\ 
\hhline{|>{\arrayrulecolor[rgb]{0.925,0.957,1}}->{\arrayrulecolor{black}}-------|}
\rowcolor[rgb]{0.882,0.941,0.882} \multirow{-2}{*}{{\cellcolor[rgb]{0.925,0.957,1}}\textbf{Eval~}} & {\cellcolor[rgb]{0.925,0.957,1}}\textbf{F1~}  & 0.99~       & {\cellcolor[rgb]{1,1,0.78}}0.95~       & 0.96~       & {\cellcolor[rgb]{1,1,0.78}}0.99~       & 0.99~       & {\cellcolor[rgb]{1,1,0.78}}0.99~        \\
\hline
\end{tabular}
\vspace{-0.3 cm}
\end{table}
\begin{table}
\centering
\caption{Performance of \emph{spoofing attack attribution} with spoof CM embedding (${\bf e}_{\text{cm}}$) and probabilistic attribute embedding (${\bf p}_a$). Eval*: only the utterances belonging to A16 and A19 of evaluation set are considered and tested against A04 and A06 attacks, respectively. }
\label{tab2_new}
\arrayrulecolor{black}
\begin{tabular}{|c|c|c|c|c|c|c|c|} 
\hline
\rowcolor[rgb]{0.925,0.957,1} \multicolumn{2}{|c|}{{\cellcolor[rgb]{0.925,0.957,1}}}                                                                & \multicolumn{6}{c|}{\textbf{Spoofing Attack Attribution~}}                                                                                                                \\ 
\hhline{|>{\arrayrulecolor[rgb]{0.925,0.957,1}}-->{\arrayrulecolor{black}}------|}
\rowcolor[rgb]{0.925,0.957,1} \multicolumn{2}{|c|}{{\cellcolor[rgb]{0.925,0.957,1}}}                                                                & \multicolumn{2}{c|}{\textbf{AASIST~}}                      & \multicolumn{2}{c|}{\textbf{RB-AASIST~}}             & \multicolumn{2}{c|}{\textbf{SSL-AASIST~}}             \\ 
\hhline{|>{\arrayrulecolor[rgb]{0.925,0.957,1}}-->{\arrayrulecolor{black}}------|}
\rowcolor[rgb]{0.882,0.941,0.882} \multicolumn{2}{|c|}{\multirow{-3}{*}{{\cellcolor[rgb]{0.925,0.957,1}}~}}                                         & ${\bf e}_{\text{cm}}$   & {\cellcolor[rgb]{1,1,0.78}}${\bf p}_a$    & ${\bf e}_{\text{cm}}$ & {\cellcolor[rgb]{1,1,0.78}}${\bf p}_a$ & ${\bf e}_{\text{cm}}$ & {\cellcolor[rgb]{1,1,0.78}}${\bf p}_a$  \\ 
\hline
\rowcolor[rgb]{0.882,0.941,0.882} {\cellcolor[rgb]{0.925,0.957,1}}                                  & {\cellcolor[rgb]{0.925,0.957,1}}\textbf{Acc~} & \textbf{91.6~} & {\cellcolor[rgb]{1,1,0.78}}\textbf{98.7}~ & 88.0~       & {\cellcolor[rgb]{1,1,0.78}}98.6~       & 71.9~       & {\cellcolor[rgb]{1,1,0.78}}89.6~        \\ 
\hhline{|>{\arrayrulecolor[rgb]{0.925,0.957,1}}->{\arrayrulecolor{black}}-------|}
\rowcolor[rgb]{0.882,0.941,0.882} \multirow{-2}{*}{{\cellcolor[rgb]{0.925,0.957,1}}\textbf{Dev~}}   & {\cellcolor[rgb]{0.925,0.957,1}}\textbf{F1~}  & 0.92~          & {\cellcolor[rgb]{1,1,0.78}}0.99~          & 0.88~       & {\cellcolor[rgb]{1,1,0.78}}0.99~       & 0.72~       & {\cellcolor[rgb]{1,1,0.78}}0.90~        \\ 
\hline
\rowcolor[rgb]{0.882,0.941,0.882} {\cellcolor[rgb]{0.925,0.957,1}}                                  & {\cellcolor[rgb]{0.925,0.957,1}}\textbf{Acc~} & \textbf{94.7~} & {\cellcolor[rgb]{1,1,0.78}}\textbf{99.2}~ & 89.9~       & {\cellcolor[rgb]{1,1,0.78}}96.4~       & 82.1~       & {\cellcolor[rgb]{1,1,0.78}}91.6~        \\ 
\hhline{|>{\arrayrulecolor[rgb]{0.925,0.957,1}}->{\arrayrulecolor{black}}-------|}
\rowcolor[rgb]{0.882,0.941,0.882} \multirow{-2}{*}{{\cellcolor[rgb]{0.925,0.957,1}}\textbf{Eval*~}} & {\cellcolor[rgb]{0.925,0.957,1}}\textbf{F1~}  & 0.97~          & {\cellcolor[rgb]{1,1,0.78}}0.99~          & 0.94~       & {\cellcolor[rgb]{1,1,0.78}}0.98~       & 0.89~       & {\cellcolor[rgb]{1,1,0.78}}0.95~        \\
\hline
\end{tabular}
\vspace{-0.2 cm}
\end{table}

\subsection{Explanability: Shapley Values}
\begin{figure}[t!]
    \centering
    \begin{minipage}{0.49\linewidth}
        \centering
        \includegraphics[width=\linewidth, height = 5.5cm]{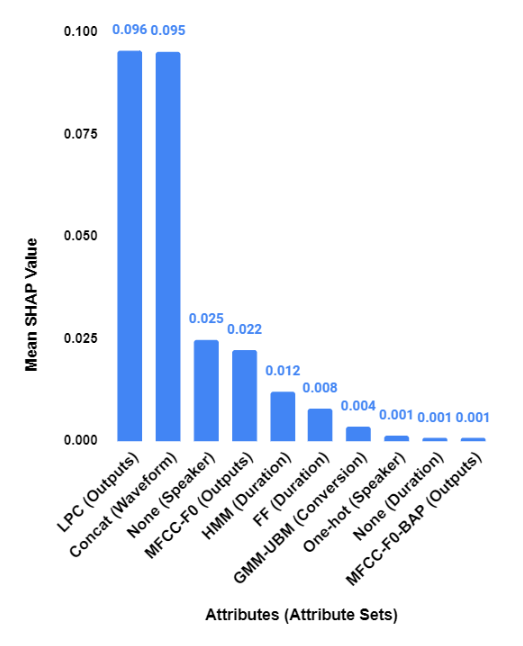}
        \subcaption{Spoofing detection}
        \label{fig:bs_shap}
    \end{minipage}
    \hfill
    \begin{minipage}{0.49\linewidth}
        \centering
        \includegraphics[width=\linewidth, height = 5.5cm]{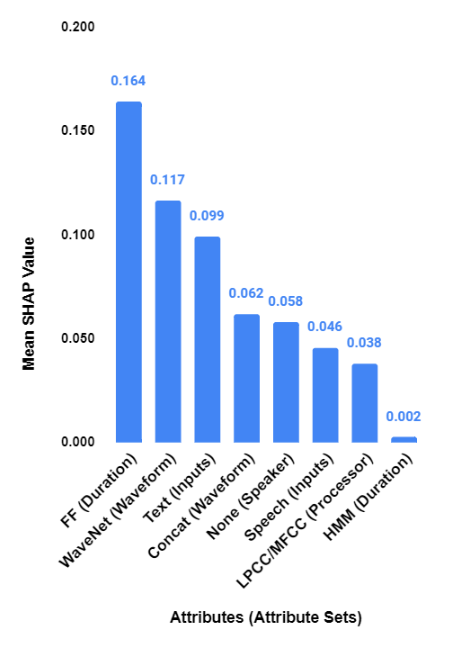}
        \subcaption{Spoofing attack attribution}
        \label{fig:saa_shap}
    \end{minipage}
    \caption{Dominating attributes with corresponding Shapley values, for spoofing detection ($\mathcal{F}^{DT}$ trained with ${\bf p}_a$ extracted from SSL-AASIST) and spoofing attack attribution ($\mathcal{F}^{DT}$ trained with ${\bf p}_a$ extracted from AASIST) tasks.}
    \label{fig:pc_combined}
    \vspace{-0.6 cm}
\end{figure}

The trained decision tree with ${\bf p}_a$ embeddings extracted from AASIST and SSL-AASIST are used to compute the feature importance through \emph{Shapley} values for spoofing detection and spoofing attack attribution, respectively.  It is observed that the tree structure is slightly varying with multiple runs with very little variation in performance. This might be due to the random splitting technique used to generate a decision tree. To mitigate the issue, the feature importance (Shapley values per attribute) is estimated by averaging the obtained feature importance in five successive runs of the decision tree. For both tasks, the feature importance is obtained in the development set. The obtained features importance for both tasks are shown in Fig.~\ref{fig:pc_combined}.



Fig.~\ref{fig:bs_shap} highlights that for spoof detection ``\texttt{LPC(Outputs)}", ``\texttt{Concat(Waveform)}", and ``\texttt{None(Speaker)}" are the most important attributes, while Fig. \ref{fig:saa_shap} indicates that ``\texttt{FF(Duration)}", ``\texttt{WaveNet(Waveform)}", and ``\texttt{Text(Inputs)}" are the most influential attributes for spoofing attack attribution. These results suggest that attributes from the modules related to acoustic feature prediction (\texttt{Outputs}), waveform generation techniques and speaker modeling play a crucial role in bonafide/spoof classification. In contrast, attributes from the duration model, waveform generation, and input type modules are notable contributors to the spoofing attack attribution task.



\section{Conclusion}

This work presents an initial investigation into the explainability of spoofed speech detection by introducing a novel approach that leverages probabilistic attributes to characterize spoofed speech. The experiments in the downstream tasks reveal the performance of the proposed probabilistic attribute embeddings is on par with the original spoof CM embeddings. Additionally, towards explainability, the Shapley values suggest that the attributes belonging to acoustic feature predictor, waveform generation, and speaker modeling are important for spoofing detection. While the attributes belonging to the duration model, waveform generation, and input type are important for spoofing attack attribution. In the future, we will attempt to generalize this framework across datasets and unknown attack types.


\bibliographystyle{plain}
\bibliography{refs}

\end{document}